\documentclass[preprint,aps]{revtex4}

\newcommand{\be}{\begin{equation}}
\newcommand{\ee}{\end{equation}}

\newcommand{\beq}{\begin{eqnarray}}
\newcommand{\eeq}{\end{eqnarray}}

\def\H1{\widehat{H}_1}

\begin{document}

\title{On the extra phase correction to the semiclassical spin coherent-state 
propagator}
\author{Mikhail Pletyukhov}
\email{pletmikh@tfp.physik.uni-karlsruhe.de}
\affiliation{Institut f\"ur Theoretische Festk\"orperphysik, Universit\"at 
Karlsruhe, D-76128 Karlsruhe, Germany}
\affiliation{Institut f\"ur Theoretische Physik, Universit\"at Regensburg, 
D-93040 Regensburg, Germany}

\begin{abstract}
The problem of an origin of the Solari-Kochetov
extra-phase contribution to the naive semiclassical form of a generalized
phase-space propagator is addressed with the special reference to the su(2)
spin case which is the most important in applications. While the
extra-phase correction to a flat phase-space propagator can straightforwardly 
be shown to appear as a difference between the principal and the Weyl symbols 
of a Hamiltonian in the next-to-leading order expansion in the semiclassical 
parameter, the same statement for the semiclassical spin coherent-state
propagator holds provided the Holstein-Primakoff representation of the su(2) 
algebra generators is employed. 
\end{abstract}

\maketitle

\section{Introduction}

The spin coherent-state path integral appears to be very useful in many
physical problems that involve quantum su(2) spins. In particular, one of its
most significant practical applications is the study of spin tunneling in the
semiclassical limit \cite{chud}-\cite{garg1}. However, as it was remarked in
\cite{stone1}, the straightforward computation of the semiclassical propagator
\cite{kur} or the tunnel splitting \cite{garg2} yields results that are
incorrect beyond the leading semiclassical order. Examples of other systems
where the large spin limit gives a good qualitative picture, while the
first quantum correction is either ignored or fixed by heuristic
considerations, can be found in \cite{stone2}.

Recently, Stone {\it et al} \cite{stone1} have restored the reliability of the
semiclassical expression for the spin coherent-state propagator, thus
effectively rehabilitating the use of the continuous-time spin coherent-state
path integral. The crucial point in that approach is the recognition of an
importance and the explanation of the origin of a previously discovered quantum
correction \cite{solari}-\cite{vieira} to the naive form of the semiclassical
coherent-state propagator. It has been pointed out that the functional
determinant arising from the evaluation of the functional integral about the
classical path possesses a U(1) gauge anomaly, its proper regularization
resulting in the extra-phase contribution. Originally Solari \cite{solari}
obtained this extra-phase correction through a careful calculation of the path
integral in the discrete-time approximation. Kochetov \cite{koch1} derived
it independently considering the continuous-time version of the spin
coherent-state path integral in the semiclassical limit. A discrete-time
evaluation similar to that of Solari was carried out by Vieira and Sacramento
\cite{vieira} who have also reproduced the same result.

The relevance of the Solari-Kochetov phase has been justified in the
application of the spin coherent-state path integral to the calculation of the
tunnel splitting of the classically degenerate ground state for a family of
models that includes a realistic approximation to the molecular magnet
${\rm Fe_8}$ \cite{stone2}. It has been also noticed in \cite{plet} that the
modification of the Gutzwiller trace formula \cite{gutz1} for systems with a
coupling of the translational and spin degrees of freedom should also contain
this extra phase in the combined limit $\hbar \to 0$, $S \to \infty$,
$\hbar S$=const.

In this paper we offer a point of view on the origin of the Solari-Kochetov
phase, which complements the analysis made in  \cite{stone1}.

To establish our notations we briefly outline in Sec. \ref{quant} some
basic facts concerning quantization on cotangent bundles and coadjoint
group orbits, the phase space manifolds that are most frequently encountered 
in applications. In Sec. \ref{fsc} we consider in detail semiclassical 
propagators on flat phase
spaces. We observe that the arising extra phase can be interpreted as a
difference between the principal and the Weyl symbols of a Hamiltonian in the
next-to-leading order in $\hbar$. On the other hand, various symbols, or
various quantization schemes, are closely related to the operator ordering
procedures. One can then alternatively state that the extra phase is the
difference between the naive classical Hamiltonian and the expectation value
of the Weyl-ordered operator \cite{stone3}. Obviously, both formulations are
equivalent, though the former provides a more efficient way to compute the
extra phases, since it is based on the Wigner-Weyl calculus and implies an
extensive use of the Moyal formula \cite{moyal}.

In Sec. \ref{semispin} we discuss the semiclassical spin coherent-state
propagator and its specific ingredient --- the Solari-Kochetov phase. We also
briefly outline the relation between the covariant and contravariant
quantization  schemes which are usually employed for the path integral
construction in the spin case.

To obtain the Solari-Kochetov phase using the analogy with the flat case we, in
the first place, need a proper definition of a Weyl symbol of the quantum spin
Hamiltonian. The point is that a standard Weyl quantization is well-defined on
a classical phase space $M$, provided it appears as a cotangent bundle to a
certain configuration space $Q$, $M=T^*Q$. While $Q$ may in principle be a
compact manifold, $T^*Q$ is always noncompact. A classical phase space of a
spin is, however, a compact finite volume manifold, the two-sphere $S^2$,
which is not a cotangent bundle. Therefore, there does not seem to exist any
natural global definition of appropriate Weyl symbols of the su(2) spin 
operators.

To partly circumvent this apparent difficulty, we employ in Sec. \ref{holstein}
a special Holstein-Primakoff representation of the su(2) algebra \cite{hp}.
It expresses the su(2) generators in terms of $\hat{a}$, $\hat{a}^{\dag}$,
$\hat{a}^{\dag} \hat{a}$ --- the standard generators of the Heisenberg-Weyl
algebra, and allows for a formal application of the standard Wigner-Weyl
calculus to spin Hamiltonians. We then are able to define the Weyl symbols of
the spin operators, provided the semiclassical representations of su(2) are
considered, and to determine their difference from the components of the
classical spin. This difference appears to be well-defined in the semiclassical
limit $S \to \infty$, exactly reproducing the Solari-Kochetov phase.

\section{\label{quant}Phase-space quantization}

The classical phase space can be thought of as a pair $(M,w)$ where $M$ is a
$2n$-dimensional Riemannian manifold and $w$ stands for a closed
non-degenerate symplectic two-form on it. A group of canonical transformations
$G$ acting on $M$ transitively leaves $w$ invariant. Quantization on $M$
amounts to constructing a complex line bundle over $M$ with a connection
one-form $\theta$ such that $d\theta =w$. A quantum Hilbert space is then
constructed out of sections of the bundle. When lifted to the bundle, the
action of $G$ gives rise to a unitary but reducible representation of $G$.
In order to select an irreducible component, the action of $G$ is restricted
to the subspace of those sections which are covariantly constant along
$n$ linearly independent vector fields $\chi_1,...,\chi_n$ on $M$
such that $w(\chi_i,\chi_j)=0$.
This procedure is usually referred to as fixing of
the polarization \cite{kirillov}.

To illustrate these definitions we briefly discuss two examples of phase
spaces, which appear to be our main concern here, cotangent bundles and
coadjoint group orbits.

Local coordinates on a cotangent bundle $M=T^*Q$ can always be separated into
``coordinates'', $q^j$, and ``momenta'', $p_j$, which are canonically
conjugated to the $q^j$, with the globally defined canonical symplectic
one-form $\theta=(1/\hbar) (p^j dq_j- q^j dp_j)$.
Covariantly constant sections $f(q,p)$
along the vector field $\chi=\partial_p$ (we put for simplicity $n=1$)
are then defined by
\be \nabla_p f:=(\partial_p+i\theta_p)f =0,
\ee
whence $f(q,p)\sim
e^{ipq/\hbar}\psi(q)$. The quantum Hilbert space appears then as the space of
square-integrable functions $\psi(q)$. When considering the time evolution, one
usually picks up initial and final states as the eigenstates of the operator
$\hat q$, $\hat q|q_0\rangle=q_0|q_0\rangle$, which corresponds to the choice
$\psi(q)\sim \delta(q-q_0)$. Analogously, the choice $\chi=\partial_q$
results in quantization in terms of the
$p$-dependent sections, $f(q,p)\sim e^{-ipq/\hbar}\psi(p)$. In the following we
will be considering quantum propagators on $T^*Q$ having the form
\be
K(p_f , q_i , T):=\langle p_f|e^{-i \hat{H}
T/\hbar}|q_i\rangle, \label{pfs}
\ee
where initial and final quantum states
belong to different polarizations.

Formally, the Weyl symbol of the operator $\hat{H} (\hat{q}, \hat{p})$ is given
by
\be H_W (q, p) = \int_{Q} dx \sqrt{g}e^{i p x/\hbar} \langle q - x/2 |
\hat{H} | q + x/2 \rangle,
\label{wey}
\ee
where $g$ denotes the determinant of a restriction of a full
metric tensor onto $Q$. As already said, the separation of the local
coordinates into $(q,p)$ is defined on a cotangent bundle globally (because
$\theta$ exists globally). Accordingly, the symbol $H_W (q,p)$ is defined on
the phase space manifold $T^*Q$ globally as well.

The second important class of classical phase spaces is the so-called coadjoint
orbits of Lie groups \cite{kirillov}. These symplectic $G$-homogeneous
manifolds appear as natural classical phase spaces in the instances where
quantum Hamiltonians allow for representations in terms of the Lie group
generators. For example, the quantum spin dynamics is governed by a Hamiltonian
built out of the su(2) generators. Spin classical phase space appears then as
an SU(2) orbit --- the two-sphere $S^2$ \cite{stone4}, which is a compact 
curved manifold. It can be covered by two local charts with complex coordinates
$(z,\bar{z})$ and $(w,\bar{w})$ defined by projections from the North and South
Poles, respectively.

Since there is no globally defined symplectic one-form on $S^2$, it does not
admit a global separation of coordinates into $q$'s and $p$'s. Instead, the
locally defined su(2) symplectic one-forms (in the representation with spin
$S$) $\theta_1 = i S (\bar{z} dz - z d \bar{z})/(1+|z|^2 )$ and $\theta_2 =i S
(\bar{w} dw - w d \bar{w})/(1+|w|^2 )$, where $w=1/z$, are related in the
charts' overlap by a U(1) gauge transformation, $\theta_1=\theta_2+i S d\log
(z/\bar{z})$.

The covariantly constant section of the monopole $P(S^2,U(1))$ bundle that
forms a quantum Hilbert space for spin can be chosen as an su(2) coherent
state. In particular,
\be
\nabla_{\bar z} |z\rangle_1 := (\partial_{\bar z}+i{\theta_1}_{\bar z})
|z\rangle_1 =0,
\ee
where $|z\rangle_1$ denotes the su(2) coherent state in the  local chart
$(z, \bar{z})$.  The su(2) quantum phase-space propagator can then be written
down in the form
\be
K(\bar{z}_f , z_i , T):=
_1\!\!\langle z_f | e^{-i \hat{H} T/\hbar} |z_i \rangle_1,
\label{prspin}
\ee
where the initial and final states belong to different polarizations.

Given a spin quantum Hamiltonian $\hat H$, its classical symbol may be
conveniently chosen as an expectation value in the coherent state, the
so-called covariant symbol of the spin-dependent operator,
$H^{cov}_1 (\bar{z},z)=
_1\!\!\langle z|\hat H|z\rangle_1$.  In spite of its
appearance, this symbol is defined globally on $S^2$: the $P(S^2,U(1))$ local
sections are related in the charts' overlap by the U(1) gauge transformation
\be
_2\langle 1/z|z\rangle_1=(z/\bar z)^S\in U(1).
\ee
Therefore, $H^{cov}_1(\bar z,z)=H^{cov}_2(\bar w,w)$ for $w=1/z$.

Concluding the section, we may state: though it might be in principle possible
to construct a Weyl symbol for the su(2) spin locally on $S^2$, there seems
to be no way
to extend such a definition globally to the whole sphere.

\section{\label{fsc}Flat phase-space propagator}

The well-known van Vleck representation  for the semiclassical propagator on a
flat configuration space \cite{vvl} has its counterpart in the phase space
$T^*R^2$ which is the semiclassical approximation for the propagator
(\ref{pfs})
\be
K (p_f , q_i , T) = \frac{1}{\sqrt{2 \pi \hbar}} \left( -\frac{\partial^2
{\cal R}_{cl}}{\partial p_f \partial q_i} \right)^{1/2}  e^{i
{\cal R}_{cl} (p_f , q_i , T)/ \hbar} ,
\label{van}
\ee
where
\be
{\cal R}_{cl} (p_f , q_i , T) = - \frac12 \left( p_f q_{cl} (T) + p_{cl} (0)
q_i \right) + \int_0^T dt \left( -\frac12(\dot{p}_{cl}  q_{cl} -
 p_{cl} \dot{q}_{cl}) - H (q_{cl} ,p_{cl}) \right) .
\ee
The classical dynamics of the system is governed by the principal symbol
$H(q,p)$ of the quantum Hamiltonian $\hat{H}$:
\beq
\dot{q} &=& \,\,\,\,\, \frac{\partial H(q,p)}{\partial p}, \quad \, q(0)=q_i ,
 \\
\dot{p} &=& - \frac{\partial H(q,p)}{\partial q}, \quad p(T)=p_f.
\eeq
We would also call $H(q,p)$ the naive classical Hamiltonian since it can be
obtained by just ``dropping hats'' in $\hat{H}$.

Formally, the principal symbol $H(q,p)$ is obtained in the limit $\hbar \to 0$
of the Weyl symbol (\ref{wey}). The Weyl symbol of the product of two operators
is given by the Moyal product \cite{moyal} of two respective symbols. The Moyal
formula can be written in the symbolic form
\be
( \hat{F} \hat{G} )_W = F_W * G_W =  F_W \, e^{i \hbar {\cal L} /2} \, G_W ,
\label{inmoy}
\ee
where the operator
\be
{\cal L} = \frac{\stackrel{\gets}{\partial}}{\partial q}
\frac{\stackrel{\to}{\partial}}{\partial p} -
\frac{\stackrel{\gets}{\partial}}{\partial p}
\frac{\stackrel{\to}{\partial}}{\partial q}
\ee
is associated with the Poisson bracket
\be
F_W {\cal L} G_W = \{ F_W , G_W \}_{q,p} = \frac{\partial F_W }{\partial q}
\frac{\partial G_W }{\partial p} - \frac{\partial F_W }{\partial p}
\frac{\partial G_W }{\partial q} .
\label{poibr}
\ee
Expanding (\ref{inmoy}) into a series of $\hbar$, we obtain the leading and
the next-to-leading contributions to the Moyal formula
\be
(\hat{F} \hat{G})_W = F_W G_W + \frac{i \hbar}{2} \{ F_W , G_W \}_{q,p} +
O (\hbar^2) .
\label{moyal}
\ee
Respectively, the leading contribution to the Weyl symbol of the commutator
\be
([\hat{F}, \hat{G}])_W = (\hat{F} \hat{G} - \hat{G} \hat{F})_W =
F_W * G_W - G_W * F_W = i \hbar \{ F_W , G_W \}_{q,p} + O (\hbar^3)
\label{corr}
\ee
establishes the ``correspondence principle'' between commutators and Poisson
brackets.

However, the expression (\ref{van}) is, in general, incorrect. Besides the
Morse index that accounts for the number of conjugate points \cite{gutz2},
there may appear another discrepancy between the representation of (\ref{van})
and that of a correct asymptotic propagator. It happens when the quantum
Hamiltonian contains the terms that mix $\hat{q}$ and $\hat{p}$. Unless
$\hat{H}$ is Weyl (symmetrically) ordered, it is not sufficient to take into
account just the principal symbol $H(q,p)$ for the calculation of ${\cal
R}_{cl}$.

Thus, we make here the following proposition: in the semiclassical
evaluation of the propagator of type (\ref{van}) it is necessary to use
the Weyl symbol of the quantum Hamiltonian:
\be
H_W (q,p) = H (q,p) - \delta H (q,p) + O (\hbar^2), \quad \delta H (q,p)
\sim O(\hbar).
\ee
Though the classical dynamics is governed by the  principal symbol, or naive
classical Hamiltonian, $H (q,p)$, the next order term $\delta H (q,p)$ divided
by $\hbar$, nevertheless, contributes to the phase of the semiclassical
propagator. If the quantum Hamiltonian is a polynomial in $\hat{q}$ and
$\hat{p}$, it is easy to calculate $\delta H (q,p)$ using the Moyal formula
(\ref{moyal}) and the obvious fact that $(\hat{q})_W =q$ and $(\hat{p})_W =p$.

Let us introduce the complex variables
\be
\alpha = \frac{1}{\sqrt{2}} \left( q + i p \right), \quad \bar{\alpha} =
\frac{1}{\sqrt{2}} \left( q - i p \right) .
\label{alp}
\ee
They are normalized to be the Weyl symbols of the operators
\be
\hat{A} = \frac{1}{\sqrt{2}} \left( \hat{q} + i \hat{p} \right) , \quad
\hat{A}^{\dag} =  \frac{1}{\sqrt{2}} \left( \hat{q} - i \hat{p} \right)
\label{biga}
\ee
which satisfy the commutation relation
\be
[ \hat{A} , \hat{A}^{\dag} ] = \hbar.
\label{aahb}
\ee

We can define the Weyl symbols $F_W (\bar{\alpha} , \alpha)$ and
$G_W (\bar{\alpha} , \alpha)$ and the Poisson bracket
\be
\{ F_W (\bar{\alpha} , \alpha), G_W (\bar{\alpha} , \alpha)\}_{\bar{\alpha} ,
 \alpha} = i \left(\frac{\partial F_W }{\partial \bar{\alpha}}
\frac{\partial G_W }{\partial \alpha} - \frac{\partial F_W }{\partial \alpha}
\frac{\partial G_W }{\partial \bar{\alpha}} \right)
\label{poia}
\ee
by making a change of variables (\ref{alp}) in $F_W (q,p)$ and $G_W (q,p)$ and
in the Poisson bracket (\ref{poibr}), respectively. In turn, after the
transformation (\ref{biga}) the operator which is Weyl-ordered in
$\hat{A}, \hat{A}^{\dag}$ converts into the operator Weyl-ordered in $\hat{q}
, \hat{p}$. Note also that for the Weyl-ordered operator (either in
$\hat{A}, \hat{A}^{\dag}$ or in $\hat{q} , \hat{p}$), its Weyl symbol yields
the principal symbol, the higher order terms being identically zero. This
justifies the above definition of the Weyl symbols
$F_W (\bar{\alpha} , \alpha)$ and $G_W (\bar{\alpha} , \alpha)$.

Taking into account a definite correspondence between the symbols and the
ordering procedures, we can therefore interpret the emergence of the extra 
phase $\delta H / \hbar$ as an artifact of the operator ordering.

For a more detailed explanation, we would like to discuss the results of
\cite{koch2}. In particular,  there has been established the relation between
semiclassical results for a propagator obtained within different quantization
schemes. The  ``$\lambda$-quantization'' infers the choice of the
$\lambda$-symbol \cite{berezin1}
\be
H^{(\lambda )} (\bar{\alpha}, \alpha) = {\rm Tr} [ \hat{H} \hat{R}_{\lambda}
(\bar{\alpha}, \alpha) ], \quad \lambda \in [0,1],
\label{lsym}
\ee
where
\be
\hat{R}_{\lambda} (\bar{\alpha}, \alpha) = \frac{1}{\pi \hbar} \int d^2 \xi
e^{-(1-2 \lambda ) \bar{\xi} \xi / 2 \hbar} e^{\{ \xi (\bar{\alpha}
- \hat{A}^{\dag}) - \bar{\xi} (\alpha - \hat{A}) \} / \hbar} .
\ee
The particular cases of $\lambda = 0, 1/2,1$ correspond to the covariant
(coherent-state), Weyl and contravariant symbols, respectively. The arbitrary
$\lambda$-symbol is linked to  the covariant symbol through the relation
\be
H^{(0)} (\bar{\alpha} , \alpha) = (\hat{T}_{\lambda} (\Delta ) H^{(\lambda)})
(\bar{\alpha} , \alpha), \quad \hat{T}_{\lambda} (\Delta ) = e^{\hbar \lambda
\Delta} ,
\label{cvct}
\ee
where $\Delta =  \partial^2 / \partial \alpha \partial \bar{\alpha} $ is the
Laplace-Beltrami operator on the complex plane.

In the semiclassical limit $\hbar \to 0$ the expression (\ref{cvct}) can be
expanded as
\be
H^{(0)} = H^{(\lambda)} + \hbar \lambda \Delta H^{(\lambda)} + O (\hbar^2) .
\label{dif1}
\ee
Therefore, for arbitrary $\lambda, \lambda' \in [0,1]$ we have
\be
H^{(\lambda')} = H^{(\lambda)} - \hbar (\lambda' - \lambda)
\Delta H^{(\lambda)} + O (\hbar^2) .
\label{dif2}
\ee
The expression for the semiclassical propagator reads \cite{koch2}
\be
K_{scl}^{flat} = \left( i
\frac{\partial^2 {\cal R}^{(\lambda)}_{cl}}{\partial \bar{\alpha}_f \partial
\alpha_i } \right)^{1/2} \exp \left\{ i
\frac{{\cal R}^{(\lambda)}_{cl}}{\hbar} + i \left( \frac12 - \lambda \right)
\int_0^T B^{(\lambda)} dt \right\} ,
\label{flat}
\ee
where
\beq
{\cal R}^{(\lambda)} (\bar{\alpha}_f , \alpha_i , T) &=& -\frac{i}{2} \left(
\bar{\alpha}_f \alpha_{cl} (T) + \bar{\alpha}_{cl} (0) \alpha_i
- | \alpha_f |^2 - | \alpha_i |^2 \right)  \nonumber \\
&+& \int_0^T dt \left( - \frac{i}{2} (\dot{\bar{\alpha}}_{cl} \alpha_{cl} -
\bar{\alpha}_{cl} \dot{\alpha}_{cl})
- H^{(\lambda )} (\bar{\alpha}_{cl} , \alpha_{cl}) \right)
\eeq
and
\be
B^{(\lambda)} = \Delta H^{(\lambda)} = \Delta H^{(1/2)} + O(\hbar)=
\Delta H^{(0)} + O(\hbar) .
\ee
The terms of the order $O(\hbar)$ in $B^{(\lambda)}$ as well as the dependence
 of the prefactor on $\lambda$ are inessential due to the very structure of the
 asymptotic expression (\ref{flat}). $O(\hbar)$-terms are also negligible in
the classical equations of motion:
\beq
\dot{\alpha} &=& - i \frac{\partial H^{(\lambda )}}{\partial \bar{\alpha}}
+ O (\hbar), \quad \, \alpha (0) = \alpha_i , \\
\dot{\bar{\alpha}} &=& \,\,\,\,\, i  \frac{\partial H^{(\lambda )}}{\partial
\alpha} + O (\hbar), \quad \bar{\alpha} (T) = \bar{\alpha}_f .
\eeq

Note that for $\lambda = 1/2$ (Weyl quantization) the  $B$-term drops out from
(\ref{flat}). Since the semiclassical propagator should not depend on
$\lambda$, i.e. on the choice of the quantization scheme, the extra-phase
correction just compensates for the difference between $\lambda$ and Weyl
symbols in the next-to-leading order in $\hbar$ [see (\ref{dif2}) for
$\lambda' = 1/2$].

Suppose that the quantum Hamiltonian $\hat{H}$ belongs to a family of
specifically ordered Hamiltonians \cite{glaub}, also parametrized by
$\lambda \in [0,1]$,
\be
\hat{H}_{\lambda} (\hat{A}^{\dag},\hat{A} )= \frac{1}{(\pi \hbar)^2} \int d^2
\alpha d^2 \beta H (\bar{\alpha} , \alpha) e^{(1-2 \lambda ) \bar{\beta} \beta
/ 2 \hbar} e^{\{ \beta (\bar{\alpha} - \hat{A}^{\dag}) - \bar{\beta} (\alpha -
\hat{A})\}  /\hbar},
\label{lord}
\ee
where the particular cases of $\lambda = 0, 1/2,1$ correspond to the normal,
Weyl and antinormal orderings, respectively. One can establish the one-to-one
correspondence between the operators (\ref{lord}) and the symbols (\ref{lsym}).
It follows from the observation that  the $\lambda$-symbol of the
$\lambda$-ordered operator yields the principal symbol
$H (\bar{\alpha} , \alpha)$. Thus, the result (\ref{flat}) of \cite{koch2}
actually proves our proposition for the Hamiltonians (\ref{lord}).

The problem of the extra-phase contribution was often encountered in the
coherent-state semiclassics based on the covariant quantization scheme.
We can illustrate it with the following example.

Let us consider the  normally ordered operator
\be
\hat{H}_{\lambda = 0} = \sum_{m,n} h_{mn} \hat{A}^{\dag m} \hat{A}^{n} .
\label{norm}
\ee

Its covariant (coherent-state) symbol
\be
H^{(0)} (\bar{\alpha} , \alpha) = \langle \alpha | \hat{H}_0 | \alpha \rangle =
\sum_{m,n} h_{mn} \bar{\alpha}^m \alpha^{n}
\label{covn}
\ee
is found from (\ref{lsym}) with $\lambda = 0$, or, equivalently, by taking an
expectation value in the Heisenberg-Weyl coherent state
\be
| \alpha \rangle =  e^{- \bar{\alpha} \alpha /2 \hbar}
e^{\alpha \hat{A}^{\dag} / \hbar}| 0 \rangle , \quad
\langle \alpha | \alpha \rangle = 1.
\ee
Obviously, the covariant symbol (\ref{covn}) coincides with the principal
symbol.

The Weyl symbol of (\ref{norm}) is
\be
H^{(1/2)} = \sum_{m,n} h_{mn} \bar{\alpha}^m \alpha^{n} + \frac{i \hbar}{2}
\sum_{m,n} h_{mn} \{ \bar{\alpha}^m , \alpha^n \}_{\bar{\alpha} , \alpha}
+ O(\hbar^2)  = H^{(0)} - \frac{\hbar}{2} \Delta H^{(0)} + O(\hbar^2) .
\ee

Thus, we can deduce that
\be
\frac{\delta H}{\hbar} = \frac{1}{2} \Delta H^{(0)} .
\label{df}
\ee
The same expression for the phase correction has been derived in 
Ref. \cite{voros}, where the WKB 
approximation in the Bargmann representation has been considered.  
In the case $h_{11} = \omega$ and $h_{mn} = 0$ for $m,n \not= 1$, we see that
$\delta H /\hbar = \omega /2 $. When multiplied by $T$, it exactly coincides
with the required phase correction to the semiclassical coherent-state
propagator of the harmonic oscillator. Its inclusion is sometimes referred to
as a restoration of zero-point energy in the first order of $\hbar$.

\section{\label{semispin}Semiclassical spin coherent-state propagator}

A quantum spin Hamiltonian $\hat{H} = \hat{H} ( {\bf \hat{s}})$ is a function
of spin algebra generators which satisfy the su(2) commutation relations
\be
[ \hat{s}_{+} , \hat{s}_{-} ] = 2 \hat{s}_{3} , \quad [ \hat{s}_{3} ,
\hat{s}_{\pm} ] = \pm \hat{s}_{\pm} ,
\label{su2}
\ee
where $\hat{s}_{\pm} = \hat{s}_{1} \pm i \hat{s}_{2}$.

There actually exist two quantization schemes for spin \cite{berezin2} ---
covariant and contravariant ---  that are usually employed for the 
path-integral  construction and its further semiclassical approximations.

The covariant quantization scheme is based on the coherent-state representation
of the quantum Hamiltonian: the covariant symbol is defined as an expectation
value
\be
H^{cov} (\bar{z} , z) \equiv H (\bar{z} , z) = _1\!\!\langle z | \hat{H} |
z \rangle_1
\label{scs}
\ee
in the spin coherent state
\be
| z \rangle_1 = (1 + \bar z z)^{-S} e^{z \hat{s}_{+}} |S, -S \rangle , \quad
_1\langle z | z \rangle_1 = 1,
\ee
where $| S , -S \rangle$ is the lowest spin state in the $2S+1$-dimensional
representation of SU(2).
The second family of the spin coherent states is given by
\be
| w \rangle_2 = (1 + \bar w w)^{-S} e^{w \hat{s}_{-}} |S, S \rangle , \quad
_2\langle w | w \rangle_2 = 1,
\ee
$|S, S \rangle$ being the highest spin state.

In the sequel, all spin coherent states will be drawn from the first family,
$|z\rangle:=|z\rangle_1$.

The spin coherent-state propagator (\ref{prspin}) can be approximated in the
semiclassical limit $S \to \infty$ by
\be
K_{scl} (\bar{z}_f , z_i , T)=
\left( i \frac{(1 + \bar{z}_f z_{cl} (T))(1 + \bar{z}_{cl} (0) z_i )}{2S}
\frac{\partial^2 {\cal R}_{cl}}{\partial z_i \partial \bar{z}_f} \right)^{1/2}
e^{ i {\cal R}_{cl} (\bar{z}_f , z_i , T) +
\frac{i}{2} \int_0^T \phi_{SK} (t) dt} .
\label{spp}
\ee
The  validity of this formula has been proven in \cite{stone1}. It  has been
also shown that the degree of its accuracy, assuming errors of at most
$O(1/S)$,  is uniform in $T$.

The leading contribution to the phase of $K_{scl}$ is the classical action
\beq
{\cal R}_{cl} ( \bar{z}_f , z_i , T) &=&
- i S \{ \ln [(1 + \bar{z}_f z_{cl} (T))(1 + \bar{z}_{cl} (0) z_i )]
- \ln [(1 + |z_f |^2)(1 + |z_i |^2)] \} \nonumber \\
&+& \int_0^T \left\{ - i S \frac{\dot{\bar{z}}_{cl} z_{cl}
- \bar{z}_{cl} \dot{z}_{cl}}{1 + \bar{z}_{cl} z_{cl}}
- H (\bar{z}_{cl} , z_{cl}) \right\} dt .
\label{clact}
\eeq
(Note the distinction  up to a factor of $i$ in our notation and that
of \cite{stone1} as well as the difference in the normalization of the spin
coherent states.) Classical trajectories $z_{cl} (t), \bar{z}_{cl} (t)$ are to
be found from the classical equations of motion
\beq
\dot{z} &=& -i \frac{(1 + \bar{z} z)^2}{2S}
\frac{\partial H}{\partial \bar{z}} , \quad \, z(0)=z_i , \\
\dot{\bar{z}} &=& \,\,\,\,\, i \frac{(1 + \bar{z} z)^2}{2S}
\frac{\partial H}{\partial z}  , \quad \bar{z} (T)=z_f .
\label{eqz}
\eeq

The Solari-Kochetov phase, or the first quantum phase correction to $K_{scl}$,
is expressed through
\be
\phi_{SK} (t) = \frac12 \left( \frac{\partial}{\partial \bar{z}}
\frac{(1+ \bar{z} z)^2}{2S} \frac{\partial H}{\partial z} +
\frac{\partial}{\partial z} \frac{(1+ \bar{z} z)^2}{2S}
\frac{\partial H}{\partial \bar{z}}\right) \biggr|_{z=z_{cl},
\bar{z} = \bar{z}_{cl}}
\label{sk}
\ee
and represents the main subject of our discussion.

In particular, for a Hamiltonian linear in the spin operators
\be
\hat{H} = {\bf C} \cdot \hat{{\bf s}} = \frac12  C_{+} \hat{s}_{-} +
\frac12 C_{-} \hat{s}_{+} + C_3 \hat{s}_3 ,
\ee
where $C_{\pm} = C_1 \pm i C_2$, we obtain the spin coherent-state symbol 
\be
H (\bar{z} , z) = S \frac{C_{+} z + C_{-} \bar{z}}{1 + \bar{z} z}-
S C_3 \frac{1 - \bar{z} z}{1 + \bar{z} z} ,
\label{linz}
\ee
and the Solari-Kochetov phase correction
\be
\frac12 \phi_{SK} = -\frac14 (C_{+} z_{cl} + C_{-} \bar{z}_{cl})
+ \frac{1}{2} C_3 .
\label{skz}
\ee

Another example is the Hamiltonian of the Lipkin-Meshkov-Glick (LMG) model 
\cite{lmg}
\be
\hat{H}_{LMG} = \frac{w}{\sqrt{2} (2S-1)} (\hat{s}^2_{+} + \hat{s}^2_{-}) + 
\frac{S w}{\sqrt{2}},
\label{lip}
\ee
which is quadratic in the spin operators. In this case we have 
\be
H_{LMG}  (\bar{z}, z) = \sqrt{2} S w \frac{\bar{z}^2 + z^2}{(1 + \bar{z} z)^2}
 + \frac{S w}{\sqrt{2}}
\ee
and
\be
\frac12 \phi_{SK}^{LMG} = - \frac{w}{\sqrt{2}} 
\frac{(\bar{z}^2 + z^2 )(2 + \bar{z} z)}{(1 + \bar{z} z)^2}.
\ee
The latter has been calculated in \cite{stone2}, and its
inclusion  into the semiclassical propagator has provided the correct result 
for the tunnel splitting of the ground state in the LMG model.

To complete the presentation of the semiclassical spin propagator, we would
also like to mention another quantization scheme which relies on the
contravariant symbol $H^{ctr} (\bar{z} , z)$ given by
\be
\hat{H} = \frac{2S+1}{\pi}
\int \frac{d^2 z}{(1+ \bar{z} z)^2} H^{ctr} (\bar{z}, z) |z \rangle \langle z|.
\ee
As follows from~\cite{berezin2},
the relation between the covariant and contravariant symbols in the
semiclassical limit $S \to \infty$ reads
\be
H^{cov} (\bar{z} , z) = [1 + \Delta + O (1/S^2)] H^{ctr} (\bar{z} , z),
\label{ctr}
\ee
where
\be
\Delta = \frac{(1+ \bar{z} z)^2}{2S} \frac{\partial^2}{\partial z \partial
\bar{z}}
\ee
is the Laplace-Beltrami operator acting on the complex projective plane
$CP^{1}=S^2$ which is a K\"ahler homogeneous manifold
SU(2)/U(1). In view of (\ref{ctr}), one may convert formula (\ref{spp}) into 
a form suitable for the quantization by contravariant symbols. 

\section{\label{holstein}Solari-Kochetov phase from Hol\-stein-Pri\-ma\-koff 
representation}

Now we would like to derive the Solari-Kochetov phase exploiting
the paradigm of the Sec. \ref{fsc}. For this purpose we employ the  
Holstein-Primakoff representation \cite{hp} for the spin
operators
\beq
\hat{s}_{+} &=& \hat{s}_{1} + i\hat{s}_{2} = \hat{a}^{\dag} \sqrt{2S
- \hat{a}^{\dag} \hat{a}} , \nonumber \\
\hat{s}_{-} &=& \hat{s}_{1} - i\hat{s}_{2} = \sqrt{2S
- \hat{a}^{\dag} \hat{a}} \, \hat{a}  \label{holpr} , \\
\hat{s}_3 &=& \hat{a}^{\dag} \hat{a} - S , \nonumber
\eeq
in terms of the standard annihilation and creation operators $\hat{a}$ and
$\hat{a}^{\dag}$ with the commutation relation $ [ \hat{a} , \hat{a}^{\dag} ]
=1$. It is easy to check that the operators (\ref{holpr}) satisfy the su(2)
algebra (\ref{su2}) as well as
\be
\frac12 \left(\hat{s}_{+} \hat{s}_{-} + \hat{s}_{-} \hat{s}_{+} \right)
+ \hat{s}_3^2 = S(S+1).
\ee

Besides the representation (\ref{holpr}), there actually exist another 
representations of the spin algebra in terms of bosonic operators.
For the review of their properties and applications in physics, we refer to 
\cite{bla,garb}. 

It is also worth mentioning that in Ref. \cite{belin}, where the instanton 
picture of spin tunneling in the LMG model was also considered, the sort of 
the  Holstein-Primakoff representation was used in order to obtain the correct 
ground-state energy splitting. The authors established some heuristic rule 
which, however, required an adequate interpretation (see Discussion in 
\cite{belin}). Our approach will allow to refine their prescription and to 
establish the link to the consideration of the LMG model made in \cite{stone2}.

Let us introduce the semiclassical parameter $h = 1/(2S)$ and define  $\hat{A}
= \hat{a} \sqrt{h}$ and $\hat{A}^{\dag} = \hat{a}^{\dag} \sqrt{h}$, such that
\be
[\hat{A} , \hat{A}^{\dag} ] = h .
\label{aah}
\ee
We also define the operators
\beq
\hat{S}_{+} &=& h \hat{s}_{+} = \hat{A}^{\dag}
\sqrt{1 - \hat{A}^{\dag} \hat{A}} , \nonumber \\
\hat{S}_{-} &=& h \hat{s}_{-} =
\sqrt{1 - \hat{A}^{\dag} \hat{A}} \, \hat{A} , \label{holpr1} \\
\hat{S}_3 &=& h \hat{s}_3 = \hat{A}^{\dag} \hat{A} - \frac12 , \nonumber
\eeq
which satisfy the commutation relations
\be
[ \hat{S}_{+} , \hat{S}_{-} ] = 2 h \hat{S}_3 , \quad
[ \hat{S}_3 , \hat{S}_{\pm} ] = \pm h \hat{S}_{\pm} .
\label{ncr}
\ee
The square root in (\ref{holpr1}) should be understood as an expansion in a
Taylor series
\be
\sqrt{1 - x} = 1 + \sum_{l=1}^{\infty} b_l x^l
\label{expan}
\ee
with $x$ replaced by $\hat{A}^{\dag} \hat{A}$.

One can immediately notice that the operators (\ref{holpr1}), when expressed
through $\hat{A}$ and $\hat{A}^{\dag}$, do not depend explicitly on $h$, and
that (\ref{aah}) is similar to (\ref{aahb}). This enables us to apply formally
the Moyal formula (\ref{moyal}) with the Poisson bracket (\ref{poia}) to the
operators (\ref{holpr1}), replacing everywhere $\hbar$ by $h$. Considering
$\alpha$ and $\bar{\alpha}$ to be the ``Weyl symbols'' of $\hat{A}$ and
$\hat{A}^{\dag}$, respectively, we can thus define the ``Weyl symbols'' of the
operators (\ref{holpr1}).

First we find
\beq
(\hat{A}^{\dag} \hat{A})_W &=& \bar{\alpha} \alpha - \frac{h}{2} + O(h^2), \\
(\hat{A} \hat{A}^{\dag})_W &=& \bar{\alpha} \alpha + \frac{h}{2} +O(h^2), \\
((\hat{A}^{\dag} \hat{A})^l)_W &=& (\bar{\alpha} \alpha)^l
- \frac{h l}{2} (\bar{\alpha} \alpha)^{l-1} + O(h^2) .
\eeq
Exploiting the latter relation and the trivial equality
\be
\frac{d}{d x} \sqrt{1 - x} = - \frac{1}{2 \sqrt{1 - x}}
= \sum_{l=1}^{\infty} l b_l x^{l-1}
\ee
we establish
\be
(\sqrt{1 - \hat{A}^{\dag} \hat{A}})_W = \sqrt{1 - \bar{\alpha} \alpha} +
\frac{h}{4} \frac{1}{\sqrt{1 - \bar{\alpha} \alpha}}  + O(h^2) .
\ee
Further use of the Moyal formula (\ref{moyal}) leads to the desired definitions
\beq
(\hat{S}_{+})_W &=& (\hat{A}^{\dag} \sqrt{1 - \hat{A}^{\dag} \hat{A}})_W =
\bar{\alpha} \sqrt{1 - \bar{\alpha} \alpha}+
\frac{h}{2} \frac{\bar{\alpha}}{\sqrt{1 - \bar{\alpha} \alpha}} + O(h^2),
\nonumber \\
(\hat{S}_{-})_W &=& (\sqrt{1 - \hat{A}^{\dag} \, \hat{A}} \hat{A})_W =
\alpha \sqrt{1 - \bar{\alpha} \alpha}+
\frac{h}{2} \frac{\alpha}{\sqrt{1 - \bar{\alpha} \alpha}} + O(h^2),
\label{hcor} \\
(\hat{S}_3)_W &=& (\hat{A}^{\dag} \hat{A} - \frac12)_W =
\bar{\alpha} \alpha - \frac12 - \frac{h}{2} + O(h^2) . \nonumber
\eeq

We can construct the approximate realization of the su(2) algebra
with respect to the Poisson bracket (\ref{poia}). Taking into account
(\ref{corr}) and the commutation relations (\ref{ncr}) we deduce that
\beq
i  \{ (\hat{S}_{+})_W , (\hat{S}_{-})_W \}_{\bar{\alpha} , \alpha}
&=& 2   (\hat{S}_3)_W + O(h^2) , \label{poi1} \\
i  \{ (\hat{S}_{3})_W , (\hat{S}_{\pm})_W \}_{\bar{\alpha} , \alpha}
&=&  \pm  (\hat{S}_{\pm})_W + O(h^2) . \label{poi2}
\eeq
These formulae can be checked by straightforward calculation using 
(\ref{hcor}).

There exists, however, a subtlety that should be spelled out here. The spin
operators (\ref{holpr1}) act in the finite Hilbert space in contrast to the
operators  $\hat{q}, \hat{p}$ and $\hat{H} (\hat{q}, \hat{p})$ in the flat case
which act in a different --- infinite --- Hilbert space. Nevertheless, the
``Weyl  symbols'' (\ref{hcor}) of the operators (\ref{holpr1}) do make sense
locally in the semiclassical limit $h \to 0$, and, as we shall see, reproduce
the Solari-Kochetov phase. Similarly to the flat case, we are going to
recognize it for a Hamiltonian linear in the spin operators
\be
\hat{H} = \frac{1}{h} {\bf C} \cdot \hat{{\bf S}} = \frac{1}{h} \left[ \frac12
 C_{+} \hat{S}_{-} + \frac12 C_{-} \hat{S}_{+} + C_{3} \hat{S}_3 \right]
\label{linh}
\ee
in the difference between its principal and  ``Weyl'' symbols
\be
 H (\bar{\alpha} , \alpha) - H_W (\bar{\alpha} , \alpha)
= \delta H (\bar{\alpha} , \alpha) + O(h) .
\label{spindif}
\ee
(There is a small distinction in notations in comparison with the flat case
since the classical action and the naive classical Hamiltonian for spin are
already divided by the semiclassical parameter $h$ and therefore
$H (\bar{\alpha} ,\alpha)$ and $\delta H (\bar{\alpha} , \alpha)$ are of order
$O(h^{-1})$ and $O(1)$, respectively.)

According to (\ref{hcor}) the  principal symbol and the next-order correction
of (\ref{linh}) are given, respectively, by
\beq
H (\bar{\alpha} , \alpha) &=& \frac{1}{h}
\left[ \frac12 C_{+} \alpha \sqrt{1 - \bar{\alpha} \alpha}
+ \frac12 C_{-}\bar{\alpha} \sqrt{1 - \bar{\alpha} \alpha}
+ C_3 \left(\bar{\alpha} \alpha - \frac12  \right) \right] , \label{prin} \\
\delta H (\bar{\alpha} , \alpha) &=&
-  \frac{C_{+} \alpha + C_{-} \bar{\alpha}}{4 \sqrt{1 - \bar{\alpha} \alpha}}
+ \frac{1}{2} C_3 . \label{delh}
\eeq

To compare these expressions with (\ref{linz}) and (\ref{skz}), respectively,
we use the Darboux transformation
\be
z = \frac{\alpha}{\sqrt{1 - \bar{\alpha} \alpha}}, \quad
\bar{z} = \frac{\bar{\alpha}}{\sqrt{1 - \bar{\alpha} \alpha}} .
\label{chvar}
\ee
It makes the K\"ahler symplectic structure locally flat and converts the
classical equations of motion (\ref{eqz}) into
\be
\dot{\alpha}= -i  h \frac{\partial H (\bar{\alpha} , \alpha)}{\partial
\bar{\alpha}} + O(h), \quad
\dot{\bar{\alpha}} = i h \frac{\partial H (\bar{\alpha} , \alpha)}{\partial
\alpha} + O(h).
\label{eqa}
\ee
The terms $O(h)$ appear due to the finiteness  of the $(\alpha , \bar{\alpha})$
phase space which is a disc on the complex plane. However, they become
negligible as $h \to 0$. All the other terms in (\ref{eqa}) have the order
$O(1)$.

Thus, after the transformation (\ref{chvar}) we observe the coincidence of the
principal symbols (\ref{linz}) and (\ref{prin}) and obtain the desired
relation
\be
\frac12 \phi_{SK} = \delta H .
\label{ff}
\ee

Our consideration is not restricted to the case of Hamiltonians linear in 
spin operators. The formula (\ref{ff}) can be proved valid when the 
Hamiltonian is a more general element of the enveloping algebra (i.e. a 
polynomial in spin operators). 

Let us consider, for example, the Hamiltonian
\be
\hat{H} = c \cdot\frac{(2S-n)!}{(2S-1)!}  (\hat{s}^n_{+} + \hat{s}^n_{-}),
\label{ndegspin}
\ee
where $n$ is an arbitrary integer number. For $n=2$ and $c=w/\sqrt{2}$ it 
coincides (up to the constant factor) with the Hamiltonian of the LMG model 
(\ref{lip}). 

We note that it is important to introduce  the 
$S$-dependent coefficient in (\ref{ndegspin}) so that to make the respective 
covariant symbol proportional to $S$:
\be
H^{cov} = c \cdot 2 S \frac{\bar{z}^n + z^n}{(1+ \bar{z} z)^n}.
\label{prinn}
\ee
This allows to identify (\ref{prinn}) with the principal symbol, or classical
Hamiltonian, which contains the leading-in-$S$ term only.

First, we calculate the Solari-Kochetov phase according to the original 
formula (\ref{sk}), and obtain
\be
\frac12 \phi_{SK} = - c \cdot \frac{n}{2} 
\frac{(\bar{z}^n + z^n)(n + \bar{z} z)}{(1+ \bar{z} z)^n}.
\label{solkn}
\ee

Now we would like to show that the same expression can be obtained from 
(\ref{ff}) using the Holstein-Primakoff representation. We rewrite the 
Hamiltonian (\ref{ndegspin}) in terms of the operators (\ref{holpr1})
\be
\hat{H} = c \cdot h^{-n} \frac{(h^{-1}-n)!}{(h^{-1}-1)!} 
(\hat{S}^n_{+} + \hat{S}^n_{-}),
\label{quanrewr}
\ee
and find its ``Weyl'' symbol 
\beq
(\hat{H})_W &=&  c \cdot \frac{1}{h} \left( 1+ \frac{h n (n-1)}{2} \right) 
\left[\frac{\bar{z}^n +z^n}{(1 + \bar{z} z)^n}+ \frac{h n}{2} \frac{\bar{z}^n +
 z^n}{(1 + \bar{z} z)^{n-1}}  \right] + O(h)
\nonumber \\ 
&=& c \cdot \left[ \frac{1}{h} \frac{\bar{z}^n +z^n}{(1 + \bar{z} z)^n} + 
\frac{n}{2} \frac{(\bar{z}^n +z^n)(n + \bar{z} z)}{(1 + \bar{z} z)^n}\right] + 
O (h). 
\label{wigdegn}
\eeq
This expression is derived due to
\beq
(\hat{S}^n_{+})_W &=& (\bar{\alpha} \sqrt{1 - \bar{\alpha} \alpha})^n +  
\frac{h n}{2} \bar{\alpha}^n  (\sqrt{1 - \bar{\alpha} \alpha})^{n-2} + O (h^2) 
\nonumber \\
&=& \frac{\bar{z}^n}{(1 + \bar{z} z)^n}+ \frac{h n}{2} \frac{\bar{z}^n}{(1 + 
\bar{z} z)^{n-1}} +O (h^2), \\
(\hat{S}^n_{-})_W &=& (\alpha \sqrt{1 - \bar{\alpha} \alpha})^n + 
\frac{h n}{2}   \alpha^n (\sqrt{1 - \bar{\alpha} \alpha})^{n-2} + O (h^2) 
\nonumber \\
&=& \frac{z^n}{(1 + \bar{z} z)^n} + \frac{h n}{2} \frac{z^n}{(1 + 
\bar{z} z)^{n-1}} + O (h^2), 
\eeq
and
\be
\frac{(N-n)!}{N!} = N^{-n} \left(1 + \frac{1}{N} \frac{n (n-1)}{2}  + 
O \left( \frac{1}{N^2}\right) \right),
\ee
for $N \equiv 2S = h^{-1}$. The relations (\ref{chvar}) have 
been also employed.

Thus, we see that the leading term in (\ref{wigdegn}) coincides with 
(\ref{prinn}), and the next-to-leading term is in agreement with (\ref{solkn}).
We note that in our derivation of the Solari-Kochetov phase it
was important to expand in a series of $h$ the $h$-dependent coefficient ---
the ratio of two factorials --- which was inherited from the quantum 
Hamiltonian (\ref{quanrewr}).

\section{\label{discussion}Discussion}

We put forward a proposition to determine the extra-phase correction in the
semiclassical expression for a propagator as the difference between the
principal and the Weyl symbol of the quantum Hamiltonian. Based on the
Wigner-Weyl calculus, it becomes a well-defined and efficient computational
prescription.

We offered to exploit this paradigm for the  case of the spin propagator,
making use of the Holstein-Primakoff representation of the su(2) algebra.
However, there exists a subtle issue concerning the finiteness of the spin
Hilbert space. It also shows up on the classical level: the classical phase
space in such a representation is a flat disc which has finite volume.
Nevertheless, in the semiclassical limit $S \to \infty$ the difference between
the principal and the ``Weyl'' symbols appears to be well-defined. This is also
confirmed by the possibility to construct the su(2) algebra realization in
terms of the Poisson bracket $i \{ \cdot , \cdot \}_{\bar{\alpha} , \alpha}$
with the required accuracy (modulo terms $O(h^2)$).  Obviously, this amounts to
 defining the spin Weyl symbols only locally.

We applied the developed prescription for the calculation of the phase 
correction to the semiclassical spin coherent-state propagator for systems 
with the Hamiltonian which is either linear or nonlinear in spin operators. The
presented consideration can be straightforwardly generalized for any
Hamiltonian which is a polynomial in spin operators. Our prescription is rather
 simple from a computational point of view, and it does not require the use of 
the polynomial tensor operators (cf. \cite{stone3}).

In summary, we would like to motivate the  usefulness of the ``Weyl'' symbol 
for spin with the following reference. The idea to introduce such symbol is 
employed in order to reveal the similarity between the phase 
corrections to the semiclassical flat and spin propagators, and it appears 
quite naturally when one considers  the semiclassical limit of the 
Holstein-Primakoff (HP) representation  including  the next-to-leading term. 
In Ref. \cite{belin} the  Lipkin-Meshkov-Glick (LMG) model  (quadratic in spin)
 has been considered  using the sort of the HP  representation. The 
authors recognized the necessity  of the phase  correction to the semiclassical
spin propagator within their approach and formulated the heuristic rule
for its calculation. However, they stated the lack of its adequate
interpretation. On  the other hand, the consideration of the LMG model
made in Ref. \cite{stone2} uses the original Solari-Kochetov expression for the
 phase correction. The relation  between these two approaches is missing,
although both of them  have led to the correct result.  Thus, introducing
the ``Weyl'' symbol for spin helps to bridge the gap between two different
interpretations of the same extra-phase correction.

\section*{ACKNOWLEDGEMENTS}
The author would like to thank Evgueny Kochetov for many stimulating
discussions, critical reading of the manuscript and suggestions that helped
 to improve the presentation considerably. Matthias Brack, Hajo Leschke, Simone
Warzel and Oleg Zaitsev are acknowledged for their useful comments. The work
has been supported by the Deutsche Forschunsgemeinschaft and by a travel
grant of the Heisenberg-Landau program.

\end{document}